\documentclass[twocolumn,amssymb,eqsecnum,aps,pra]{revtex4-2}
\usepackage{appendix}
\usepackage{amsmath}
\usepackage{enumerate}
\usepackage{graphicx}
\begin{document}
\def \bbeta {{\beta}\hskip -5.6pt {\beta}}
\def \brho {{\rho} \hskip -4.5pt { \rho}}
\def \bnab {{\bf\nabla}\hskip-8.8pt{\bf\nabla}\hskip-9.1pt{\bf\nabla}}
\title{Conservation of energy and momentum for an electromagnetic
field propagating into a linear medium from the vacuum}
\author{Michael E. Crenshaw}
\affiliation{US Army Combat Capabilities Development Command Aviation and Missile Center (DEVCOM AvMC), Redstone Arsenal, AL 35898, USA}
\begin{abstract}
\par
The form of the energy--momentum tensor when a quasimonochromatic
field propagates into and through an antireflection-coated, sourceless,
transparent, continuous, linear magneto-dielectric medium, initially
at rest in the local frame, remains controversial.
The Minkowski energy--momentum tensor is the main component of the
electromagnetic conservation law.
It has been known for over a century that the electromagnetic 
conservation law is unsound as evidenced by alternative
energy--momentum tensors that have been proposed to ameliorate known
physical deficiencies (violation of conservation of angular and linear 
momentum) and by the various material energy--momentum
tensors and coupling forces that have been introduced to repair or
complete the law.
The extant resolution is to treat the continuum electromagnetic system
as a subsystem and add a phenomenological material subsystem
energy--momentum tensor.
We show that the four-divergence of the total, electromagnetic plus 
material, energy--momentum tensor produces an energy continuity theorem
in which the two non-zero terms depend on different powers of the
refractive index $n$.
Then the extant resolution of the Abraham--Minkowski controversy
is self-inconsistent.
\par
\end{abstract}
\date{\today}
\maketitle
\par
\section{Introduction}
\par
A large region of otherwise empty space that completely contains, for
times of interest between an initial time $t_i$ and a final time $t_f$
($-\infty < t_i \le t \le t_f < \infty$), a finite
quasimonochromatic field propagating toward and through a sourceless,
transparent, continuous, linear magneto-dielectric medium that is 
initially at rest in the local frame is a thermodynamically
closed system $\Sigma$, definitively, regardless of
what subsystems that one chooses to identify \cite{BIObukPLA}.
In particular, any forces and whatever material motion
that is imparted by the interaction with the field are part of the
closed system along with the incident, refracted, transmitted, and
reflected fields. 
While conservation principles can be unambiguously applied
\cite{BIBrevCons} to the thermodynamically closed system, just
described, continuum electrodynamics has traditionally been
formulated solely in terms of the macroscopic Maxwell field equations 
(see Eqs.~(\ref{EQj2.01})) and the constitutive relations,
${\bf D}={\bf E}+{\bf P}=\varepsilon {\bf E}$ and
${\bf B}={\bf H}+{\bf M}=\mu {\bf H}$.
\par
The total energy--momentum tensor of the system, when the field is
in the medium, has come to be viewed as being composed of an
electromagnetic part and a material part \cite{BIMoller,BIPenHaus}.
Presenting the emerging viewpoint in 1979, Brevik \cite{BIBrev}
commented that there exists ``no unique prescription for the
separation of this total energy--momentum tensor into a field
part and a matter part.''
In their 2007 review of the Abraham--Minkowski controversy,
Pfeifer, Nieminen, Heckenberg, and Rubinsztein-Dunlop \cite{BIPfei}
present the extant viewpoint that ``any electromagnetic
energy--momentum tensor must always be accompanied by a counterpart
material energy--momentum tensor, and that the division of the total
energy--momentum tensor into these two components is entirely
arbitrary.''
If the separation is truly arbitrary, then the allowable
electromagnetic and material components of the total momentum extend
well beyond the two specific field and material pairs prescribed
by Barnett \cite{BIBarn} and Barnett and Loudon \cite{BIBarnLou}.
Arbitrary quantities cannot be verified experimentally and there are
examples, in the experimental record \cite{BIObukPLA,BIBrev,BIBarn},
of experiments that prove the Minkowski electromagnetic momentum and
experiments that prove the Abraham momentum although experiments are
unable to discriminate between the two
momentums \cite{BIObukPLA,BIBrevCons}.
\par
In this article, we derive the well-known electromagnetic
conservation law \cite{BIBrevCons,BIJackson,BIMoller,BIPfei}
\begin{equation}
\partial_{\beta} {\sf T}_M^{\alpha\beta}=0
\label{EQj1.01}
\end{equation}
as a formal (axiomatic) theorem of the Maxwell--Minkowski equations
(macroscopic Maxwell equations) and constitutive relations for the
electromagnetic field in a gradient-index antireflection-coated,
sourceless, transparent, continuous, linear magneto-dielectric
medium in the limit that the gradient Minkowski force
${\bf f}_M=(-{\bf E}^2\nabla\varepsilon -{\bf H}^2\nabla\mu)/2$
and the concomitant reflection can be neglected.
(We use the conventions of summing over repeated indices on the
same side of the equal sign, that Greek indices are elements of
$(0,1,2,3)$, and that Roman indices from the middle of the alphabet
are in $(1,2,3)$.)
\par
The electromagnetic Minkowski \cite{BIMin} energy--momentum
tensor \cite{BIJackson,BIPfei}
\begin{subequations}
\begin{equation}
{\sf T}_{M}^{\alpha\beta} \;  = \;  \left [ \begin{matrix}
\frac{1}{2}({\bf D}\cdot{\bf E}+{\bf B}\cdot{\bf H})
&({\bf E}\times{\bf H})^1
&({\bf E}\times{\bf H})^2
&({\bf E}\times{\bf H})^3
\cr
({\bf D}\times{\bf B})^1   &{\sf W}^{11}  &{\sf W}^{12} &{\sf W}^{13}
\cr
({\bf D}\times{\bf B})^2   &{\sf W}^{21}  &{\sf W}^{22}  &{\sf W}^{23}
\cr
({\bf D}\times{\bf B})^3   &{\sf W}^{31}  &{\sf W}^{32}  &{\sf W}^{33}
\cr
\end{matrix}
\right ] \; ,
\label{EQj1.02a}
\end{equation}
where
\begin{equation}
{\sf W}^{ij}= -D^iE^j-B^iH^j+
\frac{1}{2}({\bf D}\cdot{\bf E}+{\bf B}\cdot{\bf H}) \; \delta^{ij} \; ,
\label{EQj1.02b}
\end{equation}
\label{EQj1.02}
\end{subequations}
is formally derived as the main component of the electromagnetic
conservation law, Eq.~(\ref{EQj1.01}).
The electromagnetic energy
\begin{equation}
U_{em}=\int_{\Sigma} {\sf T}_{M}^{00} \; dv
=\frac{1}{2}\int_{\Sigma}
\left ({\bf D}\cdot{\bf E}+{\bf B}\cdot{\bf H}\right ) \; dv
\label{EQj1.03}
\end{equation}
and the electromagnetic Minkowski momentum 
\begin{equation}
{\bf G}_M =\frac{1}{c}\int_{\Sigma}
({\sf T}_{M}^{10},{\sf T}_{M}^{20},{\sf T}_{M}^{30}) \; dv
= \int_{\Sigma} \frac{{\bf D}\times{\bf B}}{c} \; dv
\label{EQj1.04}
\end{equation}
form a Lorentz four-vector
$(U_{em}/c,-{\bf G}_M)$ \cite{BIBrevCons,BIMoller,BIPfei,BIWang4Vec}.
The Minkowski energy--momentum tensor has a well-defined formal
relationship to the electromagnetic conservation law and to its
axioms, the Maxwell--Minkowski equations and constitutive relations.
\par
\textit{If there were no problems with the electromagnetic conservation
law then the Minkowski momentum and the Minkowski energy--momentum
tensor, which are formally derived from the Maxwell--Minkowski
equations and constitutive relations, would be settled physics.}
Instead, the Minkowski energy--momentum tensor remains enmeshed in an
ancient controversy.
\par
In 1909, Abraham \cite{BIAbr} noted that the Minkowski
energy--momentum tensor, Eq.~(\ref{EQj1.02a}), lacks transpose
symmetry indicating violation of conservation of angular momentum.
Abraham proposed a transpose symmetric energy--momentum tensor and an
accompanying linear momentum ${\bf G}_A={\bf G}_M/n^2$ for the
field in a simple linear dielectric.
Transpose symmetry is typically considered to be a necessary
characteristic of the total energy--momentum tensor and the Abraham tensor
is extensively supported in the scientific
literature \cite{BIObukPLA,BIBrev,BIBalazs}.
Nowadays, the Abraham energy--momentum tensor is often treated
as being incomplete because the Abraham linear momentum is not globally
conserved.
For now, we concentrate on the Minkowski energy--momentum tensor, which
has its own historical imperative.
\par
The lack of transpose symmetry in the Minkowski energy--momentum tensor
has come to be deemed as acceptable based on the provenance of the
electromagnetic conservation law, Eq.~(\ref{EQj1.01}), as a theorem of
the Maxwell--Minkowski equations and constitutive relations
or based on appealing to the existence of transformations that
diagonalize certain energy--momentum
tensors \cite{BIPfei,BILL}.
Alternatively, the Minkowski energy--momentum tensor is assumed to be
incomplete, but fixable with the addition of a phenomenological
material momentum energy--momentum tensor \cite{BIPfei}.
Both of these contradictory viewpoints are supported in the
current scientific literature \cite{BIPfei,BIWang4Vec,BILL}.
\par
The lack of transpose symmetry in the Minkowski
energy--momentum tensor is not the only problem with the
electromagnetic conservation law, Eq.~(\ref{EQj1.01}).
The finite quasimonochromatic field, initially in the free space
portion of the thermodynamically closed system, propagates toward
and then enters the sourceless, transparent, continuous, linear
magneto-dielectric medium at normal incidence through a gradient-index
antireflection coating.
(The quasimonochromatic field has a constant amplitude throughout
its duration except for a short smooth ramp up in amplitude at
turn-on and a short smooth ramp down in amplitude at turn-off.)
The field re-enters the vacuum through the gradient-index
antireflection coating on the opposite side of the medium.
The usual textbook \cite{BIJackson,BIGriffiths,BIZangwell,BIMar}
constitutive relations for a quasimonochromatic field
with center frequency $\omega_p$ propagating through an anti-reflection
coated, sourceless, transparent, continuous, linear magneto-dielectric
medium, at rest in the local frame,
(with dispersion treated in lowest-order \cite{BIBarnLou4})
are ${\bf D} =\varepsilon(\omega_p,{\bf r}) {\bf E} $,
${\bf B}=\mu(\omega_p,{\bf r}) {\bf H }$, and
$n=\sqrt{\varepsilon(\omega_p,{\bf r})\mu(\omega_p,{\bf r})}$.
\par
In the Fresnel drag experiments \cite{BIWeins,BIFizeau}, motion of the
medium in the local frame is a specified condition of the system and the
velocity in the local frame has an effect on the constitutive relations.
Here, the medium is initially at rest in the local frame and unless the
radiation pressure and duration are extraordinary, the 
velocity of the medium in the local frame will be quite small.
Then the effect of the material velocity on the parameters,
$\varepsilon$, $\mu$, and $n$, can be treated as negligible,
as is done in the textbook derivation of Fresnel reflection
\cite{BIJackson,BIGriffiths,BIZangwell,BIMar}.
Even more so for the current case because the gradient-index
antireflection coating makes the acceleration and velocity of the
material in the local frame negligible.
\par
While the field is inside the medium, the electric field ${\bf E}$ is
reduced in amplitude ($\mu<\varepsilon$ at optical
frequencies) compared to the incident electric field ${\bf E}_0$ in
vacuum with ${\bf E}=\sqrt{n/\varepsilon}{\bf E}_0$.
Similarly, the magnetic field ${\bf H}$ is enhanced in amplitude 
compared to the amplitude in the vacuum with
${\bf H}=\sqrt{\varepsilon/n}{\bf H}_0$.
The longitudinal width $w=w_0/n$ of the field (in the direction of
propagation) is reduced by a factor of $n$ due to the reduced speed
of light in a linear medium.
Then the linear momentum
\begin{subequations}
\begin{equation}
{\bf G}_M=\int_{\Sigma}
\left (
\varepsilon\sqrt{\frac{n}{\varepsilon}}
{\bf E}_0\times\mu\sqrt{\frac{\varepsilon}{n}}{\bf H}_0\right )
\frac{dv}{c}
\label{EQj1.05a}
\end{equation}
\begin{equation}
{\bf G}_M \;\sim \; \frac{n|{\bf E}_0||{\bf H}_0| w_0 A_0}{c}
\label{EQj1.05b}
\end{equation}
\label{EQj1.05}
\end{subequations}
that is obtained by evaluating the Minkowski linear momentum,
Eq.~(\ref{EQj1.04}), while the quasimonochromatic field, with
cross-sectional area $A_0$ is inside the medium, is a nominal
factor of $n$
\cite{BIPfei,BIWang4Vec,BIJMP,BIFofn1,BIFofn2,BITorch,BIBarnLou4}
greater than the total linear momentum of the incident field.
(The Abraham momentum ${\bf G}_A={\bf G}_M/n^2$ is a factor of $n$
smaller than the incident linear momentum.)
\par
The existence of a difference between the linear momentum of the
incident field and the Minkowski linear momentum in the medium has
been known for a long time and has been extensively documented in the
scientific literature.
In the old days, the difference in linear momentum was attributed to
the action of a Minkowski pull-force by the field entering the
medium \cite{BIWang4Vec}.
More recently, it has been attributed to a material
pseudomomentum \cite{BIGord} and a canonical material
momentum \cite{BIBarn,BIBarnLou},
\par
Nevertheless, it continues to be reported that the Minkowski linear
momentum is provably conserved \cite{BIPfei} (or `almost'
conserved \cite{BIPfei}) or that the energy, Eq.~(\ref{EQj1.03}),
and Minkowski momentum, Eq.~(\ref{EQj1.04}), form a Lorentz
four-vector \cite{BIBrevCons,BIMoller,BIPfei,BIWang4Vec}
based on the electromagnetic conservation law, Eq.~(\ref{EQj1.01}),
despite the longstanding and widespread knowledge that the Minkowski 
linear momentum is contradicted by the momentum of the incident 
field by a non-negligible factor of $n$, Eq.~(\ref{EQj1.05b})
\cite{BIPfei,BIJMP,BIFofn1,BIFofn2,BITorch,BIBarnLou4}.
The Minkowski linear momentum is not globally conserved.
\par
The electromagnetic conservation law, Eq.~(\ref{EQj1.01}), is a formal
(axiomatic) theorem of the Maxwell--Minkowski equations and constitutive
relations for an antireflection-coated, sourceless, transparent, linear
magneto-dielectric medium so that the contradiction with global
conservation of linear momentum, by a factor of $n$, as well as the
contradiction with conservation of angular momentum, explicitly
disproves the axioms.
Then it has been known for over a century, or should have been known, 
that the Maxwell--Minkowski equations are formally false.
Instead, it has been surmised that the total system is comprised of
an electromagnetic subsystem and a material subsystem.
Then the macroscopic Maxwell equations for the electromagnetic field
in a simple linear medium are incomplete and the Maxwell model is 
incomplete.
\par
Section II of this article presents the axioms and theorems that
are necessary for a rigorous derivation of the energy and momentum
continuity equations of classical continuum electrodynamics, including
the electromagnetic conservation law, as formal (axiomatic) theorems
of the Maxwell--Minkowski equations and constitutive relations.
The Minkowski tensor is explicitly identified as the energy--momentum
tensor component of the formally derived electromagnetic conservation
law.
We emphasize the formal theory aspect because if a theorem of a formal
theory is proven false then the axioms of the formal theory are proven
false.
\par
In Section III, we discuss the simultaneous conservation of energy 
and momentum.
Whereas the kinematic energy $mv^2/2$ and momentum $mv$ of a particle
are linearly independent in the dynamical variable, the energy and
linear momentum in a linear medium are both quadratic in the fields.
The kinetic energy of the material is second-order in smallness and
can be neglected in the limiting case under consideration 
such that the electromagnetic energy,
Eq.~(\ref{EQj1.03}), is equal to the total energy
\begin{equation}
U_{tot}= U_{em} \; .
\label{EQj1.06}
\end{equation}
In order to be simultaneously conserved, the total energy and the
total momentum in a linear medium must have the same dependence on
the material permeability $\varepsilon$, permittivity $\mu$, and
refractive index $n$ because the total energy and momentum are not
linearly independent. 
We find that the total momentum formula \cite{BIGord}
\begin{equation}
{\bf G}_{tot}= \int_{\Sigma} \frac{n{\bf E}\times{\bf H}}{c} dv 
\label{EQj1.07}
\end{equation}
has the same dependence on the refractive index as the
total energy, Eq.~(\ref{EQj1.06}), (Eq.~(\ref{EQj1.03})).
\par
In Sec.~IV, we review the procedure of adding a material
subsystem energy--momentum tensor to the Minkowski energy--momentum
tensor, where the latter is now viewed as an electromagnetic
subsystem tensor.
The total energy--momentum tensor ${\sf T}_{tot}^{\alpha\beta}$
and/or its main elements appear extensively in the scientific
literature \cite{BIPfei,BIBarn}.
We prove that the conservation law
$\partial_{\beta} {\sf T}_{tot}^{\alpha\beta}=0$
that is constructed from the total, electromagnetic plus material,
energy--momentum tensor is false because the four-divergence of
the total energy--momentum tensor produces an energy continuity
equation in which the two non-zero terms depend on different
powers of the refractive index $n$.
The energy continuity equation
$\partial_{\beta} {\sf T}_{tot}^{0\beta}=0$
is self-inconsistent and also violates Poynting's theorem.
\par
We conclude by reiterating that the electromagnetic conservation law,
Eq.~(\ref{EQj1.01}), is a formal theorem of the Maxwell--Minkowski
equations and constitutive relations for a quasimonochromatic 
field traversing a sourceless, transparent, linear magneto-dielectric
medium in the limit that the Minkowski force
${\bf f}_M=(-{\bf E}^2\nabla\varepsilon -{\bf H}^2\nabla\mu)/2$
on the gradient-index antireflection coating can be neglected.
We show that the incident linear momentum and the globally
conserved total linear momentum, Eq.~(\ref{EQj1.07}), contradict
the Minkowski momentum, Eq.~(\ref{EQj1.04}), and thereby disproves
the electromagnetic conservation law, Eq.~(\ref{EQj1.01}), in
which the Minkowski momentum appears.
Because the electromagnetic conservation law is proven
to be false by global conservation, the axioms of the law,
the Maxwell--Minkowski equations and constitutive relations,
are proven to be false.
\par
\section{Formal theory of the Minkowski energy--momentum tensor}
\par
The microscopic Maxwell equations for electromagnetic fields
in the vacuum of free space are fundamental.
When these laws are applied to the propagation of light in a
medium, one typically obtains the
Maxwell--Minkowski equations
\cite{BIJackson,BIGriffiths,BIZangwell,BIMar}
\begin{subequations}
\begin{equation}
\nabla\times{\bf H}- \frac{\partial {\bf D}}{\partial (ct)}=
\frac{{\bf J}_f}{c}
\label{EQj2.01a}
\end{equation}
\begin{equation}
\nabla\times{\bf E}+ \frac{\partial {\bf B}}{\partial (ct)}=0
\label{EQj2.01b}
\end{equation}
\begin{equation}
\nabla\cdot{\bf D}=\rho_f
\label{EQj2.01c}
\end{equation}
\begin{equation}
\nabla\cdot{\bf B}=0
\label{EQj2.01d}
\end{equation}
\label{EQj2.01}
\end{subequations}
for the macroscopic fields ${\bf E}$, ${\bf B}$, ${\bf D}$,
and ${\bf H}$.
Here, $\rho_f$ is the free charge density and ${\bf J}_f$ is the 
free current density.
Maxwell--Minkowski is the usual textbook
\cite{BIJackson,BIGriffiths,BIZangwell,BIMar}
representation of the macroscopic Maxwell field equations,
although there are other representations
\cite{BIKemp1,BIKemplatest,BIPenHaus,BIAnghiononi,BIIdentity}.
\par
It is straightforward to construct the electromagnetic continuity 
equations and the electromagnetic conservation law as formal
identities of the Maxwell--Minkowski equations using the rules of
algebra and calculus for scalars, vectors, matrices and tensors.
We take the scalar product of Eq.~(\ref{EQj2.01b}) with ${\bf H}$,
the scalar product of Eq.~(\ref{EQj2.01a}) with ${\bf E}$, subtract
the results, and apply a common vector identity
$\nabla\cdot({\bf X}\times{\bf Y})={\bf Y}\cdot(\nabla\times{\bf X})-
{\bf X}\cdot(\nabla\times{\bf Y})$ to produce a continuity equation
\begin{equation}
\frac{1}{c}\left (
{\bf E}\cdot\frac{\partial{\bf D}}{\partial t}
+{\bf H}\cdot \frac{\partial{\bf B}}{\partial t}
\right )
+\nabla\cdot ({\bf E}\times{\bf H})=-\frac{{\bf J}_f}{c}\cdot{\bf E} 
\label{EQj2.02}
\end{equation}
that is a formal theorem, Poynting's theorem, of the Maxwell-Minkowski
equations.
The quantity
\begin{equation}
{\bf S}= c({\bf E}\times{\bf H})
\label{EQj2.03}
\end{equation}
is identified as the Poynting energy flux vector.
\par
Adding
the vector product of ${\bf B}$ with Eq.~(\ref{EQj2.01a}),
the vector product of ${\bf D}$ with Eq.~(\ref{EQj2.01b}),
the product of Eq.~(\ref{EQj2.01d}) with $-{\bf H}$, and
the product of Eq.~(\ref{EQj2.01c}) with $-{\bf E}$
produces the well-known continuity equation \cite{BIJackson}
$$
\frac{\partial}{\partial t}\frac{{\bf D}\times{\bf B}}{c}
+{\bf D}\times(\nabla\times {\bf E})
+{\bf B}\times(\nabla\times{\bf H})
$$
\begin{equation}
-(\nabla\cdot{\bf D}){\bf E} - (\nabla\cdot{\bf B}){\bf H} =
-\rho_f{\bf E} -\frac{1}{c}{\bf J}_f\times{\bf B}
\label{EQj2.04}
\end{equation}
that is also a formal theorem of the Maxwell--Minkowski equations.
The Minkowski momentum density
\begin{equation}
{\bf g}_M= \frac{{\bf D}\times{\bf B}}{c} 
\label{EQj2.05}
\end{equation}
is identified in the first term of Eq.~(\ref{EQj2.04}).
The Minkowski momentum 
\begin{equation}
{\bf G}_M= \int_{\Sigma} \frac{{\bf D}\times{\bf B}}{c} dv
\label{EQj2.06}
\end{equation}
is the Minkowski momentum density integrated over the volume of the
system ${\Sigma}$.
The energy
\begin{equation}
U=\frac{1}{2}\int_{\Sigma}
\left ({\bf E}\cdot{\bf D}+{\bf H}\cdot{\bf B} \right ) dv
\label{EQj2.07}
\end{equation}
is the energy density
\begin{equation}
u=({\bf E}\cdot{\bf D}+{\bf H}\cdot{\bf B})/2
\label{EQj2.08}
\end{equation}
integrated over the volume.
\par
The rate at which the fields do work on a continuous distribution
of charge and current
\begin{equation}
w_{mech}={\bf J}_f\cdot{\bf E} 
\label{EQj2.09}
\end{equation}
and the Lorentz force density law
\begin{equation}
\frac{d{\bf p}_{mech}}{dt}={\bf f}_L=
\rho_f{\bf E}+ \frac{{\bf J}_f}{c}\times{\bf B} 
\label{EQj2.10}
\end{equation}
are physical interpretations of terms that are formally derived
in the energy and momentum continuity theorems, Eqs.~(\ref{EQj2.02})
and (\ref{EQj2.04}).
\par
We derived the electromagnetic energy and momentum continuity
equations as identities of the Maxwell--Minkowski equations,
Eqs.~(\ref{EQj2.01}).
In contrast, derivations of the energy and momentum continuity
equations in textbooks \cite{BIJackson,BIGriffiths,BIZangwell,BIMar}
typically start with postulating the work rate $w_{mech}$,
Eq.~(\ref{EQj2.09}), and the Lorentz force density ${\bf f}_L$,
Eq.~(\ref{EQj2.10}).
Then the postulated sources, Eqs.~(\ref{EQj2.09}) and
(\ref{EQj2.10}), are expressed in terms of the fields by
substitution of the Maxwell--Minkowski equations,
Eqs.~(\ref{EQj2.01}),
to derive the energy and momentum continuity equations,
Eqs.~(\ref{EQj2.02}) and (\ref{EQj2.04}). 
Consequently, there is scientific inertia for the presence of
the sources, $\rho_f$ and ${\bf J}_f$, and for treating continuum
electrodynamics as an open system \cite{BIBrevCons}.
\par
We seek to apply the axiomatic formal theory to a thermodynamically
closed system that consists of a large finite volume ${\Sigma}$ of
otherwise-empty space that contains, as an initial condition at time
$t_i$, the finite quasimonochromatic electromagnetic field in
the vacuum and a gradient-index antireflection-coated block of
sourceless, transparent, linear magneto-dielectric material
at rest in the local frame.
In the absence of sources and sinks the total system is
thermodynamically closed, definitively, regardless of what field
and material subsystems that one might choose to identify.
Because global conservation principles can be applied, without
ambiguity, to thermodynamically closed systems,
we specify that
\begin{subequations}
\begin{equation}
\rho_f=0
\label{EQj2.11a}
\end{equation}
\begin{equation}
{\bf J}_f=0
\label{EQj2.11b} 
\end{equation}
\begin{equation}
[\varepsilon,\; \mu,\; n=\sqrt{\varepsilon\mu}] \in \mathbb{R}\geq 1 \;,
\label{EQj2.11c} 
\end{equation}
\label{EQj2.11}
\end{subequations}
 are additional axioms of the formal theory for our sourceless
transparent system.
Then the Maxwell--Amp\`ere law, Eq.~(\ref{EQj2.01a}), and the
Gauss law, Eq.~(\ref{EQj2.01c}), become homogeneous with a
right-hand side of zero.
\par
It is straightforward to derive the homogeneous energy and momentum
continuity equations
\begin{subequations}
\begin{equation}
\frac{1}{c}\left (
{\bf E}\cdot\frac{\partial{\bf D}}{\partial t}
+{\bf H}\cdot \frac{\partial{\bf B}}{\partial t}
\right ) +\nabla\cdot ({\bf E}\times{\bf H})=0
\label{EQj2.12a}
\end{equation}
$$
\frac{\partial}{\partial t}\frac{{\bf D}\times{\bf B}}{c}
+{\bf D}\times(\nabla\times {\bf E})
+{\bf B}\times(\nabla\times{\bf H})
$$
\begin{equation}
-(\nabla\cdot{\bf D}){\bf E} - (\nabla\cdot{\bf B}){\bf H} =0 
\label{EQj2.12b}
\end{equation}
\label{EQj2.12}
\end{subequations}
as theorems of the homogeneous Maxwell--Minkowski equations
\begin{subequations}
\begin{equation}
\nabla\times{\bf H}- \frac{\partial {\bf D}}{\partial (ct)}=0
\label{EQj2.13a}
\end{equation}
\begin{equation}
\nabla\times{\bf E}+ \frac{\partial {\bf B}}{\partial (ct)}=0
\label{EQj2.13b}
\end{equation}
\begin{equation}
\nabla\cdot{\bf D}=0
\label{EQj2.13c}
\end{equation}
\begin{equation}
\nabla\cdot{\bf B}=0
\label{EQj2.13d}
\end{equation}
\label{EQj2.13}
\end{subequations}
using the same procedure that was used to derive Eqs~(\ref{EQj2.02}) and
(\ref{EQj2.04}).
\par
The usual textbook \cite{BIJackson,BIGriffiths,BIZangwell,BIMar}
derivation of the continuity laws by postulating Eqs.~(\ref{EQj2.09})
and (\ref{EQj2.10}) is not applicable to a neutral linear medium
in which $\rho_f$ and ${\bf J}_f$, and the associated work rate
$w_{mech}$ and Lorentz force density ${\bf f}_L$, do not exist.
Deriving the energy and momentum continuity theorems by formally 
combining the Maxwell equations as axioms in the manner described is
arguably more fundamental than the usual textbook procedure because
the formal axiomatic procedure works for a medium without, as well as
with, sources $\rho_f$ and ${\bf J}_f$.
\par
We define a simple linear medium as a sourceless, transparent,
isotropic, homogeneous, continuous linear magneto-dielectric medium.
Here, $\varepsilon$ is a continuum abstraction of the electric
permittivity, $\mu$ is a continuum abstraction of the
magnetic permeability, and $n$ is the macroscopic refractive index.
Dispersion is treated in lowest-order \cite{BIBarnLou4} such that
$\varepsilon$, $\mu$, and $n$ depend on the center frequency
$\omega_p$ of the incident quasimonochromatic field.
We do not include additional orders of dispersion \cite{BIBarn}
because that is an exercise in complexity for a second-order
consequence.
Unless the radiation is of extraordinary intensity and duration,
the velocity of the antireflection-coated material that is initially
at rest in the Laboratory Frame of Reference will be minimal
and neglecting the effects of the material motion on the permittivity,
permeability, and refractive index is an ``extremely accurate
approximation indeed'' \cite{BIObukPLA}.
\par
Treating the continuum electrodynamic system in lowest order,
the permittivity, permeability, and refractive index are constant
in time and a simple continuous function of location in space. 
The usual constitutive relations \cite{BIBarnLou4}, in lowest order,
\begin{subequations}
\begin{equation}
{\bf D}=\varepsilon({\omega_p,{\bf r}}) {\bf E}
\label{EQj2.14a}
\end{equation}
\begin{equation}
{\bf B}=\mu({\omega_p},{\bf r}) {\bf H}
\label{EQj2.14b}
\end{equation}
\begin{equation}
n=\sqrt{\varepsilon({\omega_p},{\bf r})\mu({\omega_p},{\bf r})} \, ,
\label{EQj2.14c}
\end{equation}
\label{EQj2.14}
\end{subequations}
where the permittivity, permeability, and refractive index are 
evaluated at the center frequency $\omega_p$ of the quasimonochromatic
field,
are additional axioms for the theoretical model of our system.
The material parameters are piecewise homogeneous (homogeneous material
and homogeneous vacuum) except for a short smooth transition modeled
as a gradient-index antireflection coating.
\par
Although it is always possible to start with a fully microscopic model
of the field and material, any microstructure of the field and
medium has been eliminated in the continuum limit.
The continuous material and continuous field cannot be reliably 
un-averaged or re-quantized at the microscopic scale once the
continuum limit is invoked.
\par
The electromagnetic continuity equations, Eqs.~(\ref{EQj2.12}),
can be written row-wise as a single differential
equation \cite{BIJackson,BIOptCommun}
using the constitutive relations, Eqs.~(\ref{EQj2.14}).
We can write Eq.~(\ref{EQj2.12b}) in component form
as \cite{BIJackson}
\begin{equation}
\frac{\partial ({\bf D}\times{\bf B})^i}{\partial (ct)}
+\sum_j\frac{\partial}{\partial x^j}{\sf W}^{ij}=
-\frac{\varepsilon{\bf E}^2}{2}\frac{\nabla_i\varepsilon}{\varepsilon}
-\frac{\mu{\bf H}^2}{2} \frac{\nabla_i \mu }{\mu} \, .
\label{EQj2.15}
\end{equation}
Then Eqs.~(\ref{EQj2.12a}) and (\ref{EQj2.15}) can be written
as a single equation \cite{BIJackson},
\begin{equation}
\partial_{\beta}{\sf T}_{M}^{\alpha\beta}= {\bf f}_M^{\alpha} \;,
\label{EQj2.16}
\end{equation}
where
\begin{subequations}
\begin{equation}
{\sf T}_{M}^{\alpha\beta} =
\left [ \begin{matrix}
\frac{1}{2}({\bf D}\cdot{\bf E}+{\bf B}\cdot{\bf H})
&({\bf E}\times{\bf H})^1
&({\bf E}\times{\bf H})^2
&({\bf E}\times{\bf H})^3
\cr
({\bf D}\times{\bf B})^1  &{\sf W}^{11}  &{\sf W}^{12} &{\sf W}^{13}
\cr
({\bf D}\times{\bf B})^2   &{\sf W}^{21}  &{\sf W}^{22}  &{\sf W}^{23}
\cr
({\bf D}\times{\bf B})^3   &{\sf W}^{31}  &{\sf W}^{32}  &{\sf W}^{33}
\cr
\end{matrix} \right ]
\label{EQj2.17a}
\end{equation}
\begin{equation}
{\sf W}^{ij}= -D^iE^j-B^iH^j+
\frac{1}{2}({\bf D}\cdot{\bf E}+{\bf B}\cdot{\bf H})\delta^{ij}
\label{EQj2.17b}
\end{equation}
\begin{equation}
{\bf f}^{\alpha}_M=
\left (0,-\frac{\varepsilon{\bf E}^2}{2}\frac{\nabla\varepsilon}
{\varepsilon}
-\frac{\mu{\bf H}^2}{2}
\frac{\nabla \mu }{\mu}\right ) 
\label{EQj2.17c}
\end{equation}
\begin{equation}
\partial_{\beta}=\left (
\frac{\partial}{\partial (ct)},
\frac{\partial}{\partial x},
\frac{\partial}{\partial y},
\frac{\partial}{\partial z}
 \right ) \;.
\label{EQj2.17d}
\end{equation}
\label{EQj2.17}
\end{subequations}
\par
Here, Eq.~(\ref{EQj2.16}) is explicitly proved to be a formal
(axiomatic) theorem of the sourceless electromagnetic continuity
equations, Eqs.~(\ref{EQj2.12}), and the constitutive relations,
Eqs.~(\ref{EQj2.14}).
Moreover, Eq.~(\ref{EQj2.16}) has been derived as a formal theorem
of the Maxwell--Minkowski equations, Eqs.~(\ref{EQj2.01}), constitutive
relations, Eqs.~(\ref{EQj2.14}), and sources (\ref{EQj2.11}), by way
of the electromagnetic continuity equations, Eqs.~(\ref{EQj2.12}).
\textit{The Minkowski energy--momentum tensor, Eq.~(\ref{EQj2.17a}),
has been formally derived for a neutral simple linear medium from the
Maxwell--Minkowski equations and constitutive relations as the main
component of the theorem, Eq.~(\ref{EQj2.16})}.
\par
In the limit that the gradient-index antireflection coating is
sufficiently smooth that we can neglect reflection, then 
the Minkowski force ${\bf f}^{\alpha}_{M}$ becomes negligible
such that Eq.~(\ref{EQj2.16}) becomes
\begin{equation}
\partial_{\beta}{\sf T}_{M}^{\alpha\beta}= 0 \;,
\label{EQj2.18}
\end{equation}
which is widely known as the electromagnetic conservation
law \cite{BIJackson}.
The vanishing right-hand side of Eq.~(\ref{EQj2.18}) proves that
there is no macroscopic force from the macroscopic electromagnetic
system acting on the medium.
The macroscopic medium remains kinematically stationary in the
local frame of reference.
Absent reflection, which is the case here, 
the electromagnetic energy and the Minkowski momentum
constitute a Lorentz four-vector $(U_{em}/c,-{\bf G}_M)$ 
\cite{BIBrevCons,BIMoller,BIPfei,BIWang4Vec}.
and the electromagnetic subsystem is thermodynamically closed.
\par
Except, the Minkowski momentum, Eq.~(\ref{EQj1.04}), that is
obtained from the Minkowski energy--momentum tensor in the 
electromagnetic conservation law for the finite quasimonochromatic
field in a linear medium is a factor of $n$ greater than the
momentum of the same field that is incident from the vacuum,
Eq.~(\ref{EQj1.05}).
Then Minkowski linear momentum is not globally conserved, by a 
non-negligible factor of $n$
\cite{BIPfei,BIWang4Vec,BIJMP,BIFofn1,BIFofn2,BITorch,BIBarnLou4}, and
the electromagnetic energy and the Minkowski momentum do not constitute
a Lorentz four-vector.
\par
\section{Energy and momentum conservation}
\par
In the continuum limit, the fields are all macroscopic fields.
The total energy and total momentum are both quadratic in the 
macroscopic fields and both must have the same dependence on the
macroscopic parameters, $\varepsilon$, $\mu$, and $n$, in order to be
simultaneously conserved.
\par
The electromagnetic fields can be written in terms of the vector
potential ${\bf A}$
as
$$
{\bf E}=-\nabla\phi-\frac{\partial {\bf A}}{\partial (ct)}
$$
$$
{\bf B}=\nabla\times{\bf A} \; .
$$
In the absence of sources, $\rho_f$ and ${\bf J}_f$, we can use
the Coulomb gauge in which $\phi=0$.
We then apply the envelope function \cite{BIBarnLou4}
\begin{equation}
{\bf A}({\bf r},t)=\frac{1}{2}\left ( 
{\bf {\tilde A}}({\bf r},t) e^{-i(\omega_p t-{\bf k}_p\cdot{\bf r})}
+c.c.\right)
\label{EQj3.01}
\end{equation}
to solve
\begin{equation}
\frac{1}{2c} \int_{\Sigma}
\left ({\bf D}\cdot{\bf E}+{\bf B}\cdot{\bf H} \right ) dv =
\int_{\Sigma}\frac{\zeta | {\bf E}\times{\bf H} | }{c} dv
\label{EQj3.02}
\end{equation}
for the unknown factor $\zeta$.
\par
We obtain the nominal factor
\begin{equation}
\zeta=n 
\label{EQj3.03}
\end{equation}
for fields with slowly varying envelopes, i.e., quasimonochromatic
fields, propagating through an anti-reflection coated, sourceless,
transparent, continuous, linear magneto-dielectric medium
that is initially at rest in the local frame.
\par
As discussed above Eq.~(\ref{EQj1.05}), the amplitudes of the electric
and magnetic fields in the medium are different from their amplitudes
in the vacuum, but the changes offset such that the product $EH$
remains constant.
Using the solution $\zeta=n$ of Eq.~(\ref{EQj3.01}), the momentum
\begin{equation}
{\bf G}_{tot}= \int_{\Sigma} \frac{n{\bf E}\times{\bf H}}{c} dv
\label{EQj3.04}
\end{equation}
is the globally conserved momentum counterpart of the globally
conserved energy for quasimonochromatic fields.
\par
The momentum density of the field inside the medium
\begin{equation}
{\bf g}_{tot}= \frac{n{\bf E}\times{\bf H}}{c}
\label{EQj3.05}
\end{equation}
is a factor of $n$ greater than the momentum density of the incident
field in the vacuum 
\cite{BIPfei,BIWang4Vec,BIJMP,BIFofn1,BIFofn2,BITorch,BIBarnLou4}.
Integrating the enhanced momentum density, Eq.~(\ref{EQj3.05}), over
the narrower pulse when the field is inside the medium proves that
the momentum, Eq.~(\ref{EQj3.04}), is globally conserved.
\par
Global conservation of the momentum, Eq.~(\ref{EQj3.04}), in the
absence of any significant reflection, was previously demonstrated
by the current author using a finite-difference time-domain numerical
solution of the wave equation with numerical integration of the
electromagnetic momentum, Eq.~(\ref{EQj3.04}), of the propagated
field at various points in time \cite{BIJMP}.
\par
Global conservation of momentum for an antireflection coated linear
medium is consistent with global conservation of energy.
The electromagnetic energy density in the linear medium,
Eq.~(\ref{EQj2.08}), is well-known to be a factor of $n$ greater than
the energy density in the vacuum.
Integrating the enhanced energy density over the narrower pulse
demonstrates global conservation of the electromagnetic energy,
Eq.~(\ref{EQj2.07}).
The enhanced energy density is due to the narrower pulse, not
to a material energy density.
Similarly, the enhanced momentum density, Eq.~(\ref{EQj3.05}), is
due to the narrower pulse, not to a material-motion momentum density.
Experiments, which are typically performed with cw fields, and the
definition of photons in the medium must take the reduced volume of
the field in the medium into account.
\par
The difference between global conservation of energy and global
conservation of momentum is that reflection changes the sign of
momentum, but not energy.
If part of the field is reflected, we must impute a momentum to the 
material that is twice the magnitude of the momentum of the reflected
field and in the direction of propagation.
Then,
\begin{subequations}
\begin{equation}
{\bf G}_{forward}=\int_{\Sigma^+} \frac{n{\bf E}\times{\bf H}}{c} dv
\label{EQj3.06a}
\end{equation}
\begin{equation}
{\bf G}_{backward}=-\int_{\Sigma^-} \frac{{\bf E}\times{\bf H}}{c} dv
\label{EQj3.06b}
\end{equation}
\begin{equation}
{\bf G}_{mat}= 2\int_{\Sigma^-} \frac{{\bf E}\times{\bf H}}{c} dv
\label{EQj3.06c}
\end{equation}
\label{EQj3.06}
\end{subequations}
where ${\Sigma^-}$ is the vacuum side of the plane of incidence ($n=1$)
and ${\Sigma^+}$ is the space on the material side of the plane
of incidence ($n$).
If any of the field has exited the medium, then there is additional
reflection and additional material momentum.
As long as the refracted field remains in the medium, the total, field
plus material, momentum is
\begin{equation}
{\bf G}_{tot}= {\bf G}_{forward}+{\bf G}_{backward}+{\bf G}_{mat} \;.
\label{EQj3.07}
\end{equation}
\par
If reflection is negligible, as is the main case here, there is no
need to assign a kinematic momentum to the material and the kinetic 
energy is nil.
The electromagnetic energy and electromagnetic momentum are
simultaneously conserved and the electromagnetic system is
thermodynamically closed such that
\begin{equation}
{\bf G}_{tot}={\bf G}_{forward}={\bf G}_{em}=
 \int_{\Sigma} \frac{n{\bf E}\times{\bf H}}{c} dv \;.
\label{EQj3.08}
\end{equation}
\par
\section{Material momentum approach}
\par
In this section, we review the basic details of the way that the
total energy--momentum tensor is presented in the scientific literature.
The extant procedure is to add a phenomenological material momentum
${\bf G}_{mat}$ to the Minkowski momentum and write the
tautology \cite{BIBarn,BIBoydMil}
\begin{equation}
{\bf G}_{tot}={\bf G}_{M}+{\bf G}_{mat}
\label{EQj4.01}
\end{equation}
such that the whole is the sum of the parts.
This is part of a larger view of the total system as consisting of an
electromagnetic subsystem and a material subsystem.
The conservation law for the total system, Ref.~\cite{BIPfei} Eq.~(10),
\begin{equation}
\partial_{\beta} {\sf T}_{tot}^{\alpha\beta}=
\partial_{\beta}\left ( {\sf T}_{M}^{\alpha\beta}
 +{\sf T}_{mat}^{\alpha\beta} \right )=0 \;,
\label{EQj4.02}
\end{equation}
is the tautological sum of the conservation law for the
electromagnetic conservation law for the electromagnetic subsystem
and the conservation law for the material motion subsystem.
In this scenario, the two subsystems are assumed to be coupled by
equal and opposite forces, $\pm {\bf f}_M$, in term of the gradients
of the permittivity and permeability.
\par
The usual model for the material is the flow of particles of dust in the
continuum limit.
In their comprehensive review article, 
Pfeifer, Nieminen, Heckenberg, and Rubinsztein-Dunlop \cite{BIPfei}
present the total, electromagnetic plus dust, energy--momentum tensor
\begin{equation}
{\sf T}_{tot}^{\alpha\beta} = \left [ \begin{matrix}
\frac{1}{2}({\bf D}\cdot{\bf E}+{\bf B}\cdot{\bf H})+\rho_0 c^2
&({\bf E}\times{\bf H}) +\rho_0 c {\bf v}
\cr
({\bf E}\times{\bf H}) +\rho_0 c {\bf v} &{\sf W}_{tot}
\cr
\end{matrix}
\right ] \; ,
\label{EQj4.03}
\end{equation}
where $\rho_0$ is the density of the dust and
${\sf W}_{tot}^{ij} ={\sf W}^{ij} +\rho_0v^iv^j$.
\par
We apply the four-divergence operator to the total energy--momentum
tensor, as shown in Eq.~(\ref{EQj4.02}).
The first row, $\partial_{\beta} {\sf T}_{tot}^{0 \beta}$, is
the energy continuity equation
\begin{equation}
\frac{1}{c}\left (
{\bf E}\cdot\frac{\partial{\bf D}}{\partial t}
+{\bf H}\cdot \frac{\partial{\bf B}}{\partial t}
\right ) +\nabla\cdot({\bf E}\times{\bf H})
+\nabla\cdot\rho_0 c {\bf v}=0 \; .
\label{EQj4.04}
\end{equation}
Pfeifer, Nieminen, Heckenberg, and Rubinsztein-Dunlop,
Ref.~\cite{BIPfei} Eq.~(44),
use global conservation of total momentum to show that
\begin{equation}
\rho_0{\bf v}=(n-1) \frac{{\bf E}\times{\bf H}}{c} \; .
\label{EQj4.05}
\end{equation}
Substituting Eq.~(\ref{EQj4.05}) into Eqs.~(\ref{EQj4.03}) and
(\ref{EQj4.04}) produces
\begin{equation}
{\sf T}_{tot}^{\alpha\beta} = \left [ \begin{matrix}
\frac{1}{2}({\bf D}\cdot{\bf E}+{\bf B}\cdot{\bf H})+\rho_0 c^2
&n({\bf E}\times{\bf H})
\cr
n({\bf E}\times{\bf H})  &{\sf W}_{tot}
\cr
\end{matrix}
\right ] \; ,
\label{EQj4.06}
\end{equation}
and
\begin{equation}
\frac{1}{c}\left (
{\bf E}\cdot\frac{\partial{\bf D}}{\partial t}
+{\bf H}\cdot \frac{\partial{\bf B}}{\partial t}
\right ) +\nabla\cdot n ({\bf E}\times{\bf H})=0 \; .
\label{EQj4.07}
\end{equation}
The total, field plus matter, momentum
${\bf G}_{tot}=\int_{\Sigma}c^{-1}
({\sf T}_{tot}^{10},{\sf T}_{tot}^{20},{\sf T}_{tot}^{20})dv$,
Eq.~(\ref{EQj3.03}), is globally conserved.
Note that the energy continuity equation, Eq.~(\ref{EQj4.04}),
obviously violates Poynting's theorem.
The next step makes it clear how much of a problem that is because
the two nonzero terms in Eq.~(\ref{EQj4.07}) depend on different
powers of $n$.
Equation~(\ref{EQj4.07}) is manifestly false.
\par
Partanen, H\"ayryen, Oksanen, and Tulkki \cite{BIPart} propose that
photons couple to the material system to form mass-polariton
quasiparticles.
Assuming that the Minkowski momentum is conserved, they write
the total, field plus matter, energy--momentum tensor as
\begin{equation}
{\sf T}_{MP}^{\alpha\beta} = \left [ \begin{matrix}
\frac{n^2}{2}
({\bf D}\cdot{\bf E}+{\bf B}\cdot{\bf H})
&n^2({\bf E}\times{\bf H})
\cr
n^2 ({\bf E}\times{\bf H})   &{\sf W}
\cr
\end{matrix}
\right ] \;.
\label{EQj4.08}
\end{equation}
In this case, the energy conservation law
$\partial_0 {\sf T}_{MP}^{\alpha 0}=0$ is self-consistent and
is apparently consistent with Poynting's theorem in a linear
medium because the common factor $n^2$ can be factored out
of the energy continuity equation 
$\partial_{\beta}{\sf T}_{MP}^{0\beta}= 0$:
\begin{equation}
\frac{n^2}{c}\left (
{\bf E}\cdot\frac{\partial{\bf D}}{\partial t}
+{\bf H}\cdot \frac{\partial{\bf B}}{\partial t}
\right ) +\nabla\cdot n^2({\bf E}\times{\bf H})=0 \; .
\label{EQj4.09}
\end{equation}
However, the mass-polariton energy--momentum tensor,
Eq.~(\ref{EQj4.08}), is false because the total, field plus matter,
momentum ${\bf G}_{MP}=\int_{\Sigma}
c^{-1}({\sf T}_{MP}^{10},{\sf T}_{MP}^{20},{\sf T}_{MP}^{20})dv$
is equal to the Minkowski momentum and is not globally conserved,
Eq.~(\ref{EQj1.05}).
The total energy ${U}_{tot}=\int_{\Sigma} {\sf T}_{MP}^{00}dv$
is not globally conserved either.
\par
The electromagnetic plus dust tensor treatment of the total
energy--momentum tensor and the mass polariton model are illustrative
of the many theories that are based on adding a material subsystem
energy--momentum tensor to the electromagnetic subsystem
energy--momentum tensor (or adding a material momentum to the
linear momentum).
However, the two subsystem energy--momentum tensors do not necessarily
add in the fashion of Eq.~(\ref{EQj4.02}) if the dynamical laws of
each subsystem are different.
This can be illustrated by considering two non-interacting systems,
one electromagnetic and one kinematic.
The two conservation laws can be combined in accordance with
Eq.~(\ref{EQj4.02}) with zero external forces on the 
two systems, but the four-divergence of the sum energy--momentum 
tensor mixes the closed systems together.
\par
\section{Conclusion}
\par
The Maxwell--Minkowski equations are not fundamental laws of physics, 
they are macroscopic equations that are phenomenologically extrapolated
from a perturbation of the fundamental microscopic Maxwell equations.
The non-negligible factor of $n$ difference between the Minkowski
linear momentum and the globally conserved total momentum in this
simplest-case thermodynamically closed system proves
that there is an essential contradiction between global
conservation of the total momentum and the electromagnetic
conservation law that is derived as an identity of the
Maxwell--Minkowski equations and constitutive relations in a
closed system that consists of a quasimonochromatic field,
initially in vacuum, incident on a gradient-index
antireflection-coated, sourceless, simple linear magneto-dielectric
material, initially at rest in the vacuum local frame.
The disproof of the electromagnetic conservation law by global 
conservation of energy and momentum in a thermodynamically closed
system proves that the axioms, the Maxwell--Minkowski equations
and constitutive relations, are formally false.
\par


\begin{thebibliography}{1}
\newcommand{\enquote}[1]{``#1''}
\par
\bibitem{BIObukPLA}
T.~Ramos, G.~F.~Rubilar and Y.~N.~Obukhov, 
\textit{Relativistic analysis of the dielectric {E}instein box: {A}braham, {M}inkowski, and the total energy--momentum tensors,}
Phys. Lett. A
\textbf{375}, 1703--1709 (2011).
\bibitem{BIBrevCons}
I. Brevik,
\textit{Radiation {F}orces and the {A}braham--{M}inkowski {P}roblem,}
Mod. Phys. Lett. A
\textbf{33}, 1830006 (2018).
\bibitem{BIMoller}
C. M\o ller,
\textit{The {T}heory of {R}elativity,}
(Oxford University Press, London, 1952).
\bibitem{BIPenHaus}
P.~Penfield, Jr. and H.~A.~Haus, 
\textit{Electrodynamics of {M}oving {M}edia}, 
MIT Press, Cambridge (1967).
\bibitem{BIBrev}
I.~Brevik, 
\textit{Experiments in phenomenological electrodynamics and the electromagnetic energy--momentum tensor,}
Phys. Rep. 
\textbf{52}, 133-201 (1979).
\bibitem{BIPfei}
R.~N.~C.~Pfeifer, T.~A.~Nieminen, N.~R.~Heckenberg, and H.~Rubinsztein-Dunlop,
\textit{Colloquium: momentum of an electromagnetic wave in dielectric media,}
Rev. Mod. Phys.
\textbf{79}, 1197--1216 (2007).
\bibitem{BIBarn}
S.~M.~Barnett, 
\textit{Resolution of the {A}braham--{M}inkowski dilemma,}
Phys. Rev. Lett. 
\textbf{104}, 070401 (2010).
\bibitem{BIBarnLou}
S.~M.~Barnett and R.~Loudon, 
\textit{The enigma of optical momentum in a medium,}
Phil. Trans. A 
\textbf{368}, 927--939 (2010).
\bibitem{BIJackson}
J.~D.~Jackson,
\textit{Classical {E}lectrodynamics,}
(John Wiley, New York, 1975).
\bibitem{BIMin}
H. Minkowski, 
\textit{Die {G}rundgleichungen f\"ur die elektromagnetischen {V}org\"ange in bewegten {K}\"orpern,}
Nachr. Ges. Wiss. G\"ott., 53-111 (1908).
\bibitem{BIWang4Vec}
C.~Wang,
\textit{Can the {A}braham {L}ight {M}omentum and {E}nergy in a {M}edium {C}onstitute a {L}orentz {F}our-{V}ector,}
J. Mod. Phys.
\textbf{4}, 1123-1132 (2013).
\bibitem{BIAbr}
M. Abraham, 
\textit{Zur {E}lektrodynamik bewegter {K}\"orper,} 
Rend. Circ. Mat. Palermo,
\textbf{28}, 1-28 (1909).
\bibitem{BIBalazs}
N.~L.~Balazs, 
\textit{The energy--momentum tensor of the electromagnetic field inside matter,}
Phys. Rev. 
\textbf{91}, 408--411 (1953).
\bibitem{BILL}
L. D. Landau and E. M. Lifshitz,
\textit{The {C}lassical {T}heory of {F}ields,}
4th ed., Elsevier, Amsterdam (2006), Sec. 32.
\bibitem{BIGriffiths}
D. J. Griffiths,
\textit{Introduction to {E}lectrodynamics,}
(Prentice Hall, New Jersey, 1981).
\bibitem{BIZangwell}
A.~Zangwill,
\textit {Modern {E}lectrodynamics,}
(Cambridge Univ. Press, Cambridge, 2011).
\bibitem{BIMar}
J.~B.~Marion and M.~A.~Heald,
\textit{Classical {E}lectromagnetic {R}adiation,} 2nd. ed.,
(Academic Press, Orlando, 1980)
\bibitem{BIBarnLou4}
S. M. Barnett and R. Loudon,
\textit{Theory of radiation pressure on magneto-dielectric materials,}
New. J. Phys. \textbf{17}, 063027 (2015).
\bibitem{BIWeins}
G. Weinstein,
\textit{Albert {E}instein and the {F}izeau 1851 {W}ater {T}ube {E}xperiment,}
arXiv:1204.3390 (2012).
\bibitem{BIFizeau}
M. H. Fizeau,
\textit{On the hypotheses relating to the luminous aether, and an experiment which appears to demonstrate that the motion of bodies alters the velocity with which light propagates itself in their interior,}
Phil. Mag. \textbf{2}, 568--573 (1851).
\bibitem{BIJMP}
M.~E.~Crenshaw,
\textit{Electromagnetic momentum and the energy-momentum tensor in a linear medium with magnetic and dielectric properties,}
J. Math. Phys.
\textbf{55}, 042901 (2014).
\bibitem{BIFofn1}
P.~L.~Saldanha and J.~S.~O.~Filho,
\textit{Hidden momentum and the {A}braham--{M}inkowski debate,}
Phys. Rev. A
\textbf{95}, 043804 (2017).
\bibitem{BIFofn2}
V.~P.~Torchigin and A.~V.~Torchigin,
\textit{Resolution of the {A}ge-{O}ld {D}ilemma about a {M}agnitude of the {M}omentum of {L}ight in {M}atter,}
Phys. Res. Int.
\textbf{2014}, 12636.
\bibitem{BITorch}
V. P. Torchigin,
\textit{Dozen arguments in favor of the {M}inkowski form of the momentum of light in matter,}
Optik \textbf{218}, 164986 (2020).
\bibitem{BIGord}
J.~P.~Gordon, 
\textit{Radiation forces and momenta in dielectric media,}
Phys. Rev. A 
\textbf{8}, 14--21 (1973).
\bibitem{BIKemp1}
B.~A.~Kemp, 
\textit{Resolution of the {A}braham--{M}inkowski debate: {I}mplications for the electromagnetic wave theory of light in matter,}
J. Appl. Phys. 
\textbf{109}, 111101 (2011).
\bibitem{BIKemplatest}
B.~A.~Kemp, 
\textit{Macroscopic {T}heory of {O}ptical {M}omentum,}
Prog. Opt. 
\textbf{20}, 437--488 (2015).
\bibitem{BIAnghiononi} 
B. Anghinoni, G. A. S. Flizikowski, L. C. Malacarne, M. Partanen, S. E. Bialkowski, and N. G. C. Astrath,
\textit{On the formulations of the electromagnetic stress-energy tensor,}
Ann. Phys. (NY) \textbf{443}, 169004 (2022).
\bibitem{BIIdentity}
M.~E.~Crenshaw,
\textit{Alternative formulation of the macroscopic field equations in a linear magneto-dielectric medium: Field equations,}
OSA Continuum
\textbf{3}, 246--252 (2020).
M.~E.~Crenshaw and T.~B.~Bahder,
\textit{Energy--momentum tensor for the electromagnetic field in a dielectric,}
Opt. Commun.
\textbf {284}, 2460--2465 (2011).
\bibitem{BIBoydMil}
P.~W.~Milonni and R.~W.~Boyd,
\textit{Momentum of light in a dielectric medium,}
Adv. Opt. Photonics
\textbf{2}, 519--553 (2010).
\bibitem{BIPart}
M.~Partanen, T.~Hayrynen, J.~Oksanen, and J.~Tulkki,
\textit{Photon mass drag and the momentum of light in a medium,}
Phys. Rev. A \textbf{95}, 063850 (2017).
\bibitem{BIOptCommun}
\end{thebibliography}
\end{document}